\documentclass[aps,twocolumn,showpacs,superscriptaddress]{revtex4}
\usepackage{amsbsy}
\usepackage{bm}
\usepackage{graphicx}

\newcommand{\nc}{\newcommand}
\nc{\da}{\dagger}
\nc{\noi}{\noindent}     
\nc{\eq}[1]{\mbox{Eq.(\ref{#1})}}
\nc{\ba}{\begin{array}}
\nc{\ea}{\end{array}}
\nc{\bea}{\begin{eqnarray}}
\nc{\eea}{\end{eqnarray}}
\nc{\fig}[1]{\mbox{Fig.~\ref{#1}}}

\nc{\bc}{\begin{center}}
\nc{\ec}{\end{center}}

\begin{document}

\title{Memory Effects in Spontaneous  Emission Processes}

\author{Arne L. Grimsmo}\email{arne.grimsmo@ntnu.no}
\affiliation{Department of Physics, 
             The Norwegian University of Science and Technology, N-7491 Trondheim, Norway}
\affiliation{Department of Physics, 
            The University of Auckland, Private Bag 92019, Auckland, New Zealand}

\author{Asle H. Vaskinn}\email{asle.vaskinn@ntnu.no}
\affiliation{Department of Physics, 
             The Norwegian University of Science and Technology, N-7491 Trondheim, Norway}
\author{Per K. Rekdal}\email{per.k.rekdal@himolde.no}
\affiliation{Molde University College, 
P.O. Box 2110,  N-6402 Molde,
Norway}

\author{Bo-Sture K. Skagerstam}\email{bo-sture.skagerstam@ntnu.no}
\affiliation{Department of Physics, 
             The Norwegian University of Science and Technology, N-7491 Trondheim, Norway}
\affiliation{Centre for Advanced Study (CAS), Drammensveien 78, N-0271 Oslo, Norway}


\begin{abstract}

We consider a quantum-mechanical analysis of spontaneous emission in terms of an effective two-level system with a vacuum decay rate $\Gamma_0$ and transition angular frequency $\omega_A$.
Our analysis is in principle exact, even though presented as a numerical solution of the time-evolution including memory effects.  The results so obtained are confronted with previous discussions in the literature. In terms of the {\it dimensionless} lifetime $\tau = t\Gamma_0$ of spontaneous emission, we 
obtain deviations from exponential decay of the form ${\cal O} (1/\tau)$ for the decay amplitude  as well as the previously obtained asymptotic behaviors of the form ${\cal O} (1/\tau^2)$ or ${\cal O} (1/\tau \ln^2\tau)$ for $\tau \gg 1 $.  The actual asymptotic behavior  depends on the adopted regularization procedure as well as on the physical parameters at hand.  We show  that for  any reasonable range of $\tau$ and for a sufficiently large value of the required angular frequency cut-off $\omega_c$ of the electro-magnetic fluctuations, i.e. $\omega_c \gg \omega_A$, one obtains  either a ${\cal O} (1/\tau)$  or a ${\cal O} (1/\tau^2)$ dependence.
In the presence of physical boundaries, which can change the decay rate with many orders of magnitude, the conclusions remains the same after a suitable rescaling of parameters.

\end{abstract}
\pacs{03.65.-w, 12.20.Ds,  32.10.-f}
\maketitle


\bc{
\section{INTRODUCTION}
\label{sec:introd}}
\ec


The development concerning the manipulation of single atoms and their interaction with the electro-magnetic field has reached an impressive state of art in recent years (see e.g. Refs. \cite{henkel_2002,henkel_2006,zimmermann_2007}).
Experimental studies have e.g.  shown that artificial atoms can lead to a Lamb shift of the order of a few per cent of the typical emission line \cite{fragner_2008}. Rather old issues concerning  the necessary deviations from exponential decay (see e.g. Refs. \cite{khaflin_58,horwitz_71,horwitz_74, exner_84,schulman_70,robiscoe_72,
sinha_72,mostowski_73,fonda_78,nakazato_94,facchi_2001}) may therefore be confronted with our current theoretical and experimental understanding of decaying quantum systems. Concerning experimental studies of deviations from the conventional exponential decay  we notice in particular the study of decaying $\tau$-leptons \cite{opal_95}, the observed  deviations in quantum-mechanical tunneling processes \cite{wilkinson_97} and the power-law behavior of the decay at times-scales larger than twenty lifetimes in dissolved organic materials \cite{minkman_2006}. In nuclear physics the decay of thorium has also been suggested as a potential target for large-time deviations \cite{tkalya_98}.

In the context elementary particle physics gauge invariant definitions of observable quantities are of central importance and the definition of a the decay width of an unstable particle is highly non-trivial (see e.g. \cite{sirlin_2001}) in this respect. In studies of a possible proton decay in Nature the time-dependence of decaying systems may also play an important role \cite{test_93}. Recently it has also been speculated that a non-exponential decay of slowly decaying ${\,}^{14}C$ nuclei may be of practical importance in radio-isotope calibration methods \cite{aston_2012}.

In the presence of material bodies with non-trivial dispersive properties, it has been observed that the decay time of atoms can be changed by many orders of magnitude (see e.g. Refs. \cite{henkel_99,rekdal_04, skagerstam_06, rekdal_07,hinds_2007}), which also has been observed in the 
laboratory \cite{jones_03,haroche_06,roux_2008,emmert_09,nouges_2009}.  A natural question to be considered could then be to what extent deviations from an exponential decay of atoms or molecules  can be observed,  a subject which has been addressed in great detail in the literature (see e.g. Refs. \cite{knight_76,cook_87,seke_88,seke_89,milonni_95}). 

In the present paper we will investigate possible deviations from exponential decay as studied in terms of an effective  two-level system interacting with a continuous number of degrees of freedom of the electro-magnetic field. Analytical as well as numerical methods will be employed and will be shown to lead to a consistent picture of the small and large time deviations from a constant decay rate.

    The paper is organized as follows. In the next section we outline the theoretical framework and explicitly verify unitarity of the two-level system.  A regularized version of the integral equation for the decay amplitude is derived in Section \ref{sec_vacuum_exact} and exact short-time and long-time expansions are obtained in Section \ref{sec_vacuum_approx}. Laplace techniques are employed in Sections \ref{sec_laplace_0} and \ref{sec_vacuum_knight_seke} in order to  compare with previous asymptotic expansion results as given in the literature and the important role of the regularization procedure made used of is emphasized.  Final remarks are given in Section \ref{sec:summary} and various technical details of some of the calculations are, for the readers convenience, summarized in Appendixes \ref{A_app}, \ref{B_app}, and \ref{C_app}.

\vspace{1.0cm}
\bc{
\section{GENERAL THEORY}
\label{sec_gen}}
\ec
\vspace{-0.5cm}

   Let us consider a neutral atom at a fixed position ${\bf r}_A$. In order to be specific, we will explicitly consider  hyper-fine interactions but the results obtained can easily be rephrased in terms of electric-dipole interactions. 
   The magnetic moment of the atom interacts with the quantized magnetic field via a conventional Zeeman coupling.
   The total, un-renormalized,  Hamiltonian has then the standard form
\bea   \nonumber
      &&~~~~~~~~~~~~ H  =      \sum_{\alpha}  \hbar \omega_{\alpha} \, | \alpha \rangle \langle \alpha | 
                   \\  \label{H}
		   &+& 
		   \int d^3r\int_{0}^{\infty} d\omega \,  \hbar \omega \, \hat{\bf f}^{\da}({\bf r},\omega) \cdot
                                                                              \hat{\bf f}({\bf r},\omega) ~ + ~ H' ~,
\eea
   \noi
   where the effective interaction part is
\bea   \label{H_prime}
  H'  &=&    -   \sum_{\alpha } \sum_{\beta} \, |\alpha \rangle \langle \beta| \; {\bm{\mu}}_{\alpha \beta}  \cdot  {\bf B}({\bf r}_A) 
		    ~~ .
\eea
  \noi
The Hamiltonian  $H$  can be regarded as a low-energy effective description of a  more fundamental and renormalizable theory of electro-magnetic processes, i.e. quantum electrodynamics.
  Here $\hat{\bf f}({\bf r},\omega)$ is an annihilation operator for the quantized magnetic field,
  $|\alpha \rangle$ denotes the atomic state and $E_{\alpha}$ is the corresponding energy.
     We assume non-degenerate states, i.e. $E_{\alpha} \neq E_{\beta}$ for $\alpha \neq \beta$.
   The magnetic moment of the atom is  ${\bm{\mu}}_{\alpha \beta} =  \langle \alpha | \hat{\bm{\mu}} | \beta \rangle$, where 
   $ \hat{\bm{\mu}}$ is the magnetic moment operator. The magnetic moment will typically be of the form ${\bm{\mu}}_{\alpha \beta}= g_S \langle \alpha|{\bf S}/\hbar|\beta \rangle e\hbar/2m_e$, where ${\bf S}/\hbar$ denotes a dimensionless spin operator and $g_S$ is an appropriate gyro-magnetic factor.
   The quantized magnetic field ${\bf B}({\bf r}) = {\bf B}^{(+)}({\bf r}) +  {\bf B}^{(-)}({\bf r})$ is expressed in terms of 
  ${\bf B}^{(+)}({\bf r}) =  \nabla \times {\bf A}^{(+)}({\bf r})$, where 
   ${\bf B}^{(-)}({\bf r}) = ( {\bf B}^{(+)}({\bf r})  )^{\da}$, and where the vector potential is given by
\bea \label{basic_field}    \nonumber
   {\bf A}^{(+)}({\bf r})  &=&     \mu_0 \, \int_0^{\infty} d \omega^{\, \prime} \int d^3 r' \, 
                                 \omega' \; \sqrt{ \frac{\hbar \epsilon_0}{\pi} \, 
                                 \epsilon_I({\bf r}',\omega') }
				 \\
				 &\times&
                    {\bf G}({\bf r},{\bf r}',\omega' ) \cdot  \hat{\bf f}({\bf r}',\omega')~ . 
\eea
   \noi
   Here the imaginary part of the complex permittivity is $\epsilon_I({\bf r},\omega)$ and obeys the Kramer-Kronig dispersion relations.
   The dyadic Green tensor ${\bf G}({\bf r},{\bf r}^{\, \prime},\omega)$ is the unique solution to the Helmholtz equation
\bea   \label{G_Helm}
  \overrightarrow{\nabla} \times \overrightarrow{\nabla} \times \bm{G}({\bf r},{\bf r}',\omega) &-& \frac{\omega ^2}{c^2}
   \epsilon({\bf r},\omega) \bm{G}({\bf r},{\bf r}',\omega) \nonumber \\ &=& \delta( {\bf r} - {\bf r}' ) \bm{1} \, , 
\eea
where the arrow in $\overrightarrow{\nabla}$ denotes a derivation with respect to the first argument in the dyadic Greens function.
Since the Helmholtz equation is a linear differential equation, the associated Green's tensor can be
   written as a sum according to
\bea \label{G_tot}
  \bf{G}({\bf r},{\bf r}',\omega)=  \bm{G}^0({\bf r},{\bf r}',\omega)
+ \bm{G}^{\textit S}({\bf r},{\bf r}',\omega) \, ,
\eea
    where $\bm{G}^0({\bf r},{\bf r}',\omega)$ represents the contribution of the direct waves from
    a point-like radiation source in an unbounded medium, which is vacuum in our case,
    and $\bm{G}^S({\bf r},{\bf r}',\omega)$ describes the scattering
    contribution of multiple  reflection waves from the body under consideration. The presence of the vacuum part
    $\bm{G}^0({\bf r},{\bf r}',\omega)$ in  Eq.(\ref{G_tot}) will in general give rise to divergences.  A regularization prescription is therefore required.  For electric-dipole transitions it is well-known that it is sufficient to subtract an energy shift, corresponding to the introduction of a renormalized mass, which leaves us with a logarithmic dependence of a cut-off frequency. 
For magnetic transitions this subtraction procedure is, however, not sufficient as will we discuss in more detail below. Our strategy is to allow for a  sufficient number of subtractions to generate a logarithmic cut-off dependence also for magnetic transitions.

   For reason of simplicity, we now limit our attention to a two-level atom approximation,
   i.e. an atom with an excited state and a ground state with the frequency transition $\omega_A \equiv (E_e - E_g)/\hbar > 0$.
   We consider the Hamiltonian in \eq{H} and apply the well-known Weisskopf-Wigner theory 
   for the transitions $e \rightarrow g$. 
   The solution to the time-dependent Schr\"odinger equation in the rotating-wave approximation (RWA), i.e. applying
   $H' \approx H_{RWA}$, where
\bea   \label{H_rwa}
   H_{RWA} = - |e \rangle \langle g|  \, {\bm{\mu}}_{eg}  \cdot  {\bf B}^{(+)}({\bf r}_A) + \textrm{h.c.} ~ ,
\eea
   \noi
   is then ($\omega_g \equiv E_g/\hbar$ and $\omega_e \equiv E_e/\hbar$)
\bea  \nonumber
  | \psi(t) \rangle \, &=& \, c_e(t) \, e^{- \, i \omega_{e}  t} \, |e \rangle \otimes |0 \rangle
                        \\  \nonumber 
			&+&
                        \int d^3r \int_0^{\infty} d \omega \;  \sum_{m=1}^3
			\\
			&\times&
			c_{g}({\bf r}, \omega,m| t) \, e^{- \, i (\omega + \omega_g) t } 
		        | g \rangle \otimes | {\bf r}, \omega, m \rangle ~ , ~~~
\label{psi_ansatz}
\eea
   \noi
   as we have ignored any higher order photon state than the $0$-photon and $1$-photon state.
   Here $|e \rangle$ and $|g \rangle$ are the excited state and ground state for the atom, respectively.
   Furthermore, $|0 \rangle$ denotes the vacuum of the electro-magnetic field and $| {\bf r}, \omega, m \rangle = \hat{f}_m^{\da}({\bf r},\omega) \, |0 \rangle$
   is a one photon state. 
 We will restrict ourselves to  the initial conditions $c_e(0)=1$ and $c_{g}({\bf r}, \omega,m| t=0)=0$.  More  general initial conditions  will require a more cumbersome analysis and will not be discussed in the present paper. The probability amplitudes $c_e(t)$ and $c_{g}({\bf r}, \omega,m| t)$ must now obey the unitarity condition
\bea   \label{unitarity}
|c_e(t)|^2 + \int d^3r \int_0^{\infty} d \omega \;  \sum_{m=1}^3 |c_{g}({\bf r}, \omega,m| t)|^2 =1\, .
\eea
   The probability amplitude for the excited atomic state $c_e(t)$ is then determined by (see e.g. Refs.\cite{scheel_2001,vogel_2006}) 
\bea   \label{cp_DL_mag}
  \frac{d{c}_e(t)}{dt} &=&  \int_0^t  dt^{\, \prime} ~ K(t-t^{\, \prime}) \, c_e(t^{\, \prime}) \, ,
\eea
    \noi
    where the un-regularized kernel $K(t)$ is
\bea
\label{K_insert}   
 K(t)  &=&    - \, \frac{1}{2 \, \pi} \;  
                                     \int_0^{\infty} d\omega \, 
                                     e^{- \, i  (\omega - \omega_A)  t  } \Gamma({\bf r}_A,\omega)~
				    \, .
\eea
   \noi
Here we have made use of 
 the conventional spin-flip decay rate for spontaneous emission as given by
\bea    \label{Gamma}
  \Gamma({\bf r}, \omega)   =
 					     \frac{2 \, \mu_0}{\hbar} \, 
					     {\bm{\mu}}_{eg} 
					     \cdot 
					     \mbox{Im}  [  \overrightarrow{\nabla} \times   
                                             {\bf G}({\bf r}, {\bf r}, \omega ) \times 
					     \overleftarrow{\nabla}  ] 
					     \cdot
					     {\bm{\mu}}_{ge} \, , ~
\eea
where the arrow in $\overleftarrow{\nabla}$ now denotes a derivation with respect to the second argument in the dyadic Greens function.
Below we will also make use of the expression
\bea   
\label{cg_DL}
   c_{g}({\bf r}, \omega,m| t)= \frac{i }{c}\omega
\sqrt{\frac{\mu_0 \varepsilon_I({\bf r},\omega)}{\pi\hbar}} \times~~~~~~~~
\nonumber \\
\left [ {\bm{\mu}}_{ge}^*\cdot \left(\overrightarrow{\nabla}
\times {\bf G}^*({\bf r}_A, {\bf r}, \omega )\right)\right]_m
I(\omega -\omega_A,t)\, ,\nonumber \\
\eea
where we have defined the integral
\bea
\label{Iomegat}
I(\omega,t )= \int_{0}^{t}dt' e^{i \omega t'} c_e(t') \, .
\eea
    \noi
The useful identity
\bea
\sum_{m=1}^3\int d^3r \frac{\omega ^2}{c^2}\varepsilon_I({\bf r},\omega)G_{jm}({\bf r}',{\bf r},\omega)G_{lm}({\bf r}'',{\bf r},\omega) = \nonumber \\ \mbox{Im}G_{jl}({\bf r}',{\bf r}'',\omega) \, , ~~~~~~~~~~~~~~~~
\eea
and Eq.(\ref{cg_DL}) now lead to
\bea
\label{cgrmt}
\int d^3r \int_0^{\infty} d \omega \;  \sum_{m=1}^3 |c_{g}({\bf r}, \omega,m| t)|^2 = \nonumber \\
 \frac{1}{2\pi}\int_0^{\infty}d\omega\, \Gamma({\bf r_A}, \omega) |I(\omega-\omega_A,t)|^2 \, .
\eea
It now follows from Eqs.(\ref{cp_DL_mag}) and (\ref{cgrmt}) that
\bea
\frac{d}{dt}\left(|c_e(t)|^2 + \int d^3r \int_0^{\infty}d \omega\sum_{m=1}^3 |c_{g}({\bf r}, \omega,m| t)|^2 \right) = 0 \, , \nonumber \\
\eea
i.e. the unitarity condition Eq.(\ref{unitarity}) is fulfilled for all times as it should. We notice, what may appear to be a trivial fact,  that  general time-dependent phase-redefinitions of the amplitudes will not change the unitary condition Eq.(\ref{unitarity}). This circumstance will, nevertheless, be useful below in order to circumvent divergent vacuum fluctuations of the theory.

  \eq{cp_DL_mag}, a well-known Volterra integral equation of second kind, may be integrated with respect to time. The result is then
\bea   \label{cp_DL}
 c_e(t) = 1   +  \int_0^t  dt' \, \kappa(t-t^{\, \prime}) \, c_e(t')  \, , 
\eea
   \noi
   where 
\bea
\kappa(t) \equiv \int_0^t  dt' \, K(t') \, ,
\eea
is the time-integrated kernel, i.e. 
\bea  \label{K}
    \kappa(t)  &=&    \frac{1}{2 \, \pi} \, 
                      \int_0^{\infty} d\omega \, 
                      \frac{e^{- \, i  (\omega - \omega_A)  t  } - 1 }{ i  (\omega - \omega_A)} \, \Gamma({\bf r}_A, \omega) \,  . ~~ 
\eea

    In passing, we mention that a rate similar to $\eq{Gamma}$ may be derived for electric-dipole transitions (see e.g. Refs.\cite{dung_2000,scheel_2001,vogel_2006,dung_2008}).
    In that case, the spontaneous decay rate $\Gamma({\bf r}, \omega)$ should be replaced by

\bea    \label{Gamma_E}
  \Gamma_E({\bf r}, \omega)   =
 					     \frac{2 \, \omega_A^2}{\hbar \, \epsilon_0 \, c^2 } \; 
					     {\bm d}_{eg} 
					     \cdot 
					     \mbox{Im}  [  {\bf G}({\bf r}, {\bf r}, \omega ) ] 
					     \cdot
					     {\bm d}_{ge} \, , ~
\eea
    \noi
    where ${\bm d}_{eg}$ is the electric dipole moment for the transition $e \rightarrow g$. By comparing Eqs.(\ref{Gamma}) and (\ref{Gamma_E}) it is now straightforward to translate between magnetic-dipole and electric-dipole transitions which will be made use of below. When referring to electric-dipole transitions we therefore assume that one make use of the appropriate
decay rate, i.e.  Eq.(\ref{Gamma_E}).

\bc{
\section{Vacuum}
\label{sec_vacuum}}
\ec
%
%
%
\bc{
\subsection{Exact}
\label{sec_vacuum_exact}}
\ec

   The dyadic Green tensor for vacuum, ${\bf G}^0({\bf r} ,  {\bf r} , \omega )$,  is given by a well-known expression (see e.g. Refs.\cite{scheel_2001,vogel_2006})
\bea     \label{G_0}
    \textrm{Im} [ \,    \overrightarrow{\nabla} \times {\bf G}^0({\bf r} ,  {\bf r} , \omega ) \times  \overleftarrow{\nabla} \, ]
                             &=&
    \frac{\omega^3}{6 \pi\, c^3} 
    \left  [ 
                                                     \ba{rrr}
                   1      &   
                   0      &   
                   0     
         \\                             
                   0      &  
                   1      &    
                   0           
         \\
                   0      & 
                   0      &  
                   1
                                                      \ea
                                                   \right ] \, , ~~~
\eea
independent of the position ${\bf r}$ as it should due to the translational invariance of the vacuum quantum state.

   \noi
   In this case the kernel in \eq{K} reduces to the  vacuum kernel $\kappa_n(t)$  and reads

\bea    \label{kappa_ins}
   \kappa_n(t)   =   \frac{\Gamma_0}{2\pi} \,   \int_0^{\omega_c} d\omega ~ (\frac{\omega}{\omega_A})^n  ~ \frac{e^{- i (\omega - \omega_A)t } - 1 }{i (\omega - \omega_A)} ~ ,
\eea
   \noi
 with $n=3$, and where we have introduced  a cut-off frequency $\omega_c$ in order to make the integral finite.
   The decay rate of magnetic spin-flip transition for a two-level atom with no angular momentum and negligible nuclear moment in free space is then (see e.g. Ref.\cite{rekdal_04})
\bea \label{gamma_0_B} 
     \Gamma_0
   = {\bar  \Gamma}_{ 0 }  S^{\, 2} \; ,
\eea 
   \noi  with
\bea
     {\bar \Gamma}_0  =  \, \mu_0  \, \frac{ ( \mu_B g_S )^2 }{3 \pi  \, \hbar} \, k_A^3 \; ,
\eea 
     \noi
     and $k_A \equiv \omega_A/c$ is the wave number in vacuum.
     Here we have introduced the dimensionless spin factor $S^{\, 2} \equiv S_x^{\, 2} + S_y^{\, 2} + S_z^{\, 2}$,
     where $S_j \equiv \langle g | \hat{S}_j/\hbar | e \rangle$ is the dimensionless matrix element component of the
     electron spin operators $\hat{S}_j$ corresponding to the transition $|e\rangle \rightarrow |g\rangle$,
     with $j=x,y,z$. Furthermore, $\mu_B = e\hbar/2m_e$ is the conventional Bohr magneton, $g_S \approx 2$ is the electron spin $g_S$ factor.

   Clearly, the time-dependent kernel in \eq{kappa_ins} is divergent as $\omega_c \rightarrow \infty$ and it is not entirely clear to us how to make sense of this kernel for all times  $t$. 
   Since, as we have mentioned above, magnetic-dipole transitions are analogues to  electric-dipole transitions, we therefore appeal to Bethe's   mass-renormalization \cite{bethe_47}  procedure as far as dealing with divergences are concerned. At large $t$ we recall that one then e.g. can make use of the distributional identity
\bea
\lim_{t \rightarrow \infty} \frac{e^{-i(\omega - \omega_A)t}-1}{\omega - \omega_A} = P\left(\frac{1}{\omega_A - \omega} \right) -i\pi\delta (\omega - \omega_A)\, , \nonumber \\
\eea
where one from the principal part $P(1/({\omega_A - \omega}))$ then adds a term $P(1/\omega)$ which, when summed over all possible final states, corresponds to Bethe's introduction of a renormalized mass. In the electric-dipole case, the finite-time kernel $\kappa_n (t)$ will then be regularized in such a manner that, for large $t$, energy shifts are reduced to at most a logarithmic dependence of the cut-off frequency $\omega_c$. As we will see below in Section  \ref{sec_vacuum_knight_seke} this regularization can be carried out in terms of  frequency shift and a conventional mass renormalization  for all times with a mass counter term.

   In the present case of magnetic transitions, which corresponds to an interaction which is not directly renormalizable in the same manner as for electric-dipole transitions (see e.g. Ref. \cite{Weinberg_95}) we, nevertheless, proceed in a manner which treat these different transitions in an equal manner as far as divergences are concerned.  We therefore subtract the second order expansion of the denominator in \eq{K} for $\omega \gg \omega_A$ in such a way that at most a  logarithmic  dependence of the cut-off frequency $\omega_c$ remains, i.e. 
\bea
  \frac{1}{\omega-\omega_A} \rightarrow
  \frac{1}{\omega-\omega_A} - \bigg ( \, \frac{1}{ \omega} + \frac{\omega_A}{\omega^2} + \frac{\omega_A^2}{\omega^3} \, \bigg ) ~ , 
\eea
  \noi
  which leaves us with the following regularized version of the kernel \eq{kappa_ins}:

\bea    \label{kappa_ins_R}
\kappa^R_0(t)   =   \frac{\Gamma_0}{2\pi} \,   \int_0^{\omega_c} d\omega ~ \frac{e^{- i (\omega - \omega_A)t } - 1 }{i (\omega - \omega_A)} ~ .
\eea
The kernel $\kappa^R_0(t)$ is therefore obtained from $\kappa_0(t)$ in Eq.(\ref{kappa_ins}) using the rule $\Gamma({\bf r}, \omega) \rightarrow \Gamma({\bf r}, \omega_A)$.
With regard to the amplitude $c_{g}({\bf r}, \omega,m| t)$ we observe that unitarity prevails if we in Eq.(\ref{cgrmt}) also make use of the same prescription. The fact that unitarity is preserved is our primary motivation for the introduction of the regularization procedure above. 
Below we will, however,  investigate the effect of keeping a dependence $\omega/\omega_A$ instead of $(\omega/\omega_A)^3$ in the kernel \eq{kappa_ins}. This would be in line with the standard  electric-dipole transition considerations as used in e.g. Refs.\cite{knight_76,cook_87,seke_88,seke_89,milonni_95}. As we will verify below, the choice of regularization procedure will effect the large time behavior of the decaying system.  

In  \eq{kappa_ins_R} we make use of  a cut-off frequency $\omega_c$ in order to make the frequency integral  finite.  Since our calculation is non-relativistic we can, e.g., identify this cut-off with
    with $m_ec^2/\hbar$,  but one may also regard this as a free parameter when one e.g. considers artificial atoms (Ref. \cite{fragner_2008} and references cited therein). We therefore introduce a dimensionless parameter cut-off parameter $\Lambda$ as defined by
\bea  \label{Lambda_cut_off}
  \Lambda \equiv \frac{\omega_c}{\omega_{A}}.
\eea
  \noi
  The subtraction procedure of $\kappa_0(t)$ above enables us to extract the leading logarithmic dependence of the cut-off parameter $\Lambda$ , and, in the end,  we have therefore replaced the vacuum kernel  $\kappa_0(t)$  with the following regularized kernel $\kappa^R_0(t)$ 
\bea \nonumber
   \kappa^R_0(t) &=&
   - \, \frac{\Gamma_0}{2\pi}  \int_{-\omega_A}^{(\Lambda - 1) \omega_A} dx \, \frac{\sin( x t ) }{x}
   \\  \label{kappa_temp}
   &-&
   i \frac{\Gamma_0}{2\pi}  \int_{\omega_A}^{(\Lambda - 1) \omega_A} dx \, \frac{\cos( x t ) }{x}
   + 
   i \frac{\Gamma_0}{2\pi} \ln (\Lambda-1) \, . ~~~~~~
\eea

   \noi
   The imaginary logarithmic term in this equation corresponds to an induced  Lamb shift due to vacuum fluctuations since  it can be removed by introducing an angular frequency shift, i.e. 
\bea   \label{omega_A_tilde}
  \tilde{\omega}_A \equiv \omega_A - \frac{\Gamma_0}{2\pi} \ln (\Lambda-1) \; ,
\eea
    \noi
    and applying the transformation
\bea \label{trans}
  c_e(t)  \rightarrow  \tilde{c}_e(t)\equiv c_e(t) \, \exp \bigg (- i \frac{\Gamma_0t}{2\pi} \, \ln (\Lambda - 1) \,  \bigg ) ~ ,
\eea
as can be seen by making use of \eq{cp_DL_mag}.
   \noi
 We also find it convenient to define the kernel 
\bea   \label{eq:kappa0} \nonumber
    \tilde{\kappa}^R_0(t) &=&    - \, \frac{\Gamma_0}{2\pi} \,  \bigg ( \, \textrm{Si}[  ({\tilde \Lambda}-1) \tilde{\omega}_A t  )  
                                                                                             +
											     \textrm{Si}[ \tilde{\omega}_A t ]  \, \bigg )
			                     \\ \label{kappa_cisi_R_0}
			                     &&  - \, i \, \frac{\Gamma_0}{2\pi} \, \bigg ( \, \textrm{Ci}[ ({\tilde \Lambda}-1) \tilde{\omega}_A t ] 
                                                                                                   - 
				                                                                   \textrm{Ci}[ \, \tilde{\omega}_A t \, ]  \, \bigg ) \, , ~~~~  
\eea
   \noi
   where $\textrm{Ci}(x)$ and $\textrm{Si}(x)$ are the standard Cosine and Sine integral, respectively (see e.g. Ref.\cite{abramowitz_70}) and ${\tilde \Lambda} \equiv \Lambda \omega_A/{\tilde \omega_A} \simeq \Lambda$ in view of the fact that $\Lambda$ is considered to be a large cut-off parameter.
The probability amplitude $\tilde{c}_e(t)$ is now a solution to the integral equation \eq{cp_DL} provided the kernel $\kappa^R_0(t)$ is replaced by $\tilde{\kappa}^R_0(t)$, i.e. we consider
\bea   \label{cp_RDL}
\tilde{c}_e(t) = 1   +  \int_0^t  dt' \, \tilde{\kappa}_0^R(t-t^{\, \prime}) \, \tilde{c}_e(t')  \, . 
\eea
The solutions of the corresponding integral equation cannot be obtained in closed form and therefore we will resort to a numerical treatment. The numerical results as presented in the present paper involves algorithms with arbitrary numerical precision.

\bc{
\subsection{Approximations}
\label{sec_vacuum_approx}}
\ec

   It is convenient to define a dimensionless time parameter $\tau \, \equiv \, \Gamma_0 t$ as well as  the parameters $b_A \equiv \omega_A/\Gamma_0$ and  $\tilde{b}_A \equiv \tilde{\omega}_A/\Gamma_0$, i.e.
\bea   \label{b_tilde}
\tilde{b}_A = b_A  - \frac{\ln (\Lambda -1)}{2\pi}  \, . 
\eea
  Let us now consider small time scales such that $({\tilde \Lambda} -1) \tilde{b}_A \tau \ll 1$ and $\tilde{b}_A \tau \ll 1$.
   Such small times are basically of academic interest as
   $t \ll \hbar/m_ec^2 = 1.29 \cdot 10^{-21} s$ unless, as suggested above,  a different scale can be provided by the use of e.g. artificial atoms. 
   Virtual particles may then be created and our theory is not strictly valid.
The leading series expansion of the kernel $\tilde{\kappa}^R_0(t)$, as given by \eq{kappa_cisi_R_0}, for such small time parameters $\tau$ is:

\bea  \label{kappa_small}
    \tilde{\kappa}^R_0(\tau) &\approx& - \, \frac{\Gamma_0}{2\pi} \, \Lambda \, \tilde{b}_A \tau 
                            -
			    \, i \, \frac{\Gamma_0}{2\pi} \, \ln ( \Lambda - 1 ) \, ,
\eea
using ${\tilde \Lambda} \approx \Lambda$. 
Apart from a sign, the imaginary part of this kernel is the same as the last imaginary part of \eq{kappa_temp}.
   It is therefore convenient to invert the transformation \eq{trans} and consider the probability amplitude $ c_e(\tau)$.
   Substituting the kernel ${\kappa}^R_0(\tau)$ into \eq{cp_DL}, we obtain
\bea \label{c_small}
   c_e(\tau) \approx    1 -  \frac{\Lambda}{4\pi \, b_A } \, ( b_A \tau )^2  \, , 
\eea
   \noi
  using the approximation $c_e(t') \approx 1$ in \eq{cp_DL} since time parameter $\tau$ is small.
   We observe that in \eq{c_small}  the parameter $b_A$ enters and not  $\tilde{b}_A$.
   In passing, we also mention that an expansion  for small times is also given in Ref.\cite{seke_88} (their Eq.(3.21)) in the case of electric-dipole transitions, with the result
\bea
\label{c_small_seke}
   c_e(\tau) \approx   1 -  \frac{\Lambda^2}{8\pi} \, ( b_A \tau )^2  \, . 
\eea
The difference between equations (\ref{c_small}) and (\ref{c_small_seke}), in which case, as mentioned above,  the factor $(\omega/\omega_A)^3$ in \eq{kappa_ins} is replaced by $\omega/\omega_A$,  illustrates the relevance of the subtraction procedure. Below we will see that this effect is even more pronounced at large vales of $\tau$.
   \noi
   The discrepancy between these last two expressions is easily explained by the fact that different regularization procedures
   have been applied.

\begin{figure}[t]
\begin{picture}(0,0)(143,255)   
\includegraphics{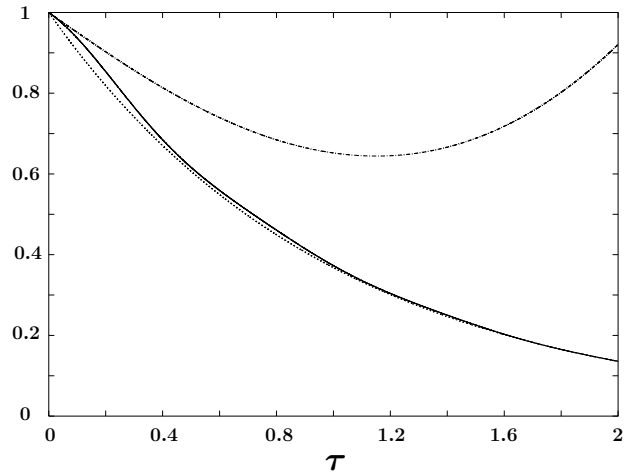}
\end{picture}
\vspace{6.4cm}
\caption{The probability $|c_e(\tau)|^2=|\tilde{c}_e(\tau)|^2$ as a function of $\tau \equiv \Gamma_0 t$.
         The solid  curve corresponds to the exact numerical solution, i.e. solution of \eq{cp_RDL} with
	 the kernel ${\tilde \kappa}_0^R(t)$ as given by \eq{kappa_cisi_R_0} in the main text for a two-level system in vacuum.
 	 The upper (dash-dotted) curve corresponds to the small time, $\tilde{b}_A \tau \ll 1$, expansion  \eq{c_e_inter}.
 	 The lower (dotted) curve corresponds to the exponential decay $\exp(-\tau)$ with expected deviations at small $\tau$.
	 The values of the relevant parameters are  $\tilde{b}_A \equiv {\tilde{\omega}}_A/\Gamma_0=10$ and $\Lambda=1000$.
}
\label{small_tau_fig}
\end{figure}

\begin{figure}[t]
\begin{picture}(0,0)(143,255)   

\includegraphics{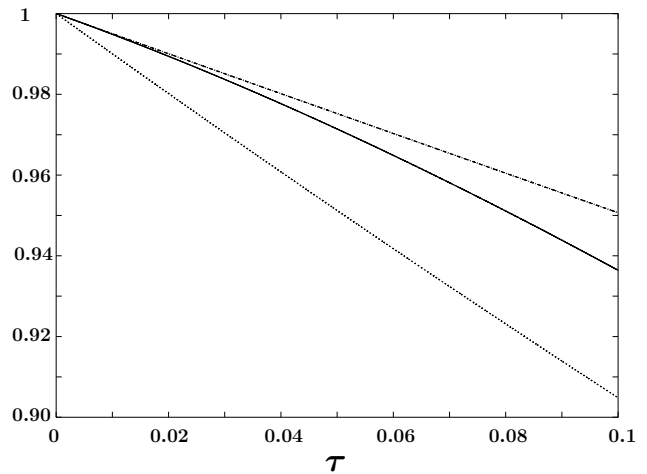}
\end{picture}
\vspace{6.4cm}
\caption{The probability $|c_e(\tau)|^2=|\tilde{c}_e(\tau)|^2$ as a function of $\tau \equiv \Gamma_0 t$
      as in \fig{small_tau_fig}, but for a  smaller $\tau$ interval.
}
\label{small_tau_fig_B}
\end{figure}

   Let us still consider small times $\tilde{b}_A \tau \ll 1$ but such that
   $({\tilde \Lambda} -1) \tilde{b}_A \tau \gg 1$. Such time scales are of relevance when one  e.g. study the Zeno  effect for quantum systems (see e.g. Refs.\cite{Zeno,facchi_2001}).  For such  intermediate times, the kernel in \eq{kappa_cisi_R_0} is reduced to
\bea   \label{k_inter}
  \tilde{\kappa}^R_0(\tau) &\approx& -  \, \frac{\Gamma_0}{4}  
			     \, + \, 
			     i \, \frac{\Gamma_0}{2\pi} \bigg (  \, \gamma_E + \ln( \tilde{b}_A \tau ) \, \bigg ) \, ,
\eea 
   where $\gamma_E \approx 0.577 \, 216$ is Euler's constant. 
   Substituting \eq{k_inter} into the \eq{cp_RDL}, expressed in terms of $\tilde{c}_e(\tau) $ and using $\tilde{c}_e(\tau') \approx 1$,  we may then carry out the time integration, with the result
\bea   \label{c_e_inter}
   \tilde{c}_e(\tau)   &\approx&     \, 1  -  \frac{\tau}{4} + i \frac{\tau}{2\pi} \bigg ( \gamma_E + \ln ( \tilde{b}_A \tau ) - 1 \, \bigg ) \,  \, .  ~~~~
\eea
   \noi
   The corresponding probability $ |\tilde{c}_e(\tau)|^2 = |c_e(\tau)|^2$
   is illustrated in Figs.\ref{small_tau_fig} and \ref {small_tau_fig_B} (dash-dotted line) together with the exact numerical solution, i.e. the numerical solution of \eq{cp_RDL} with the exact kernel
   \eq{kappa_cisi_R_0} (solid line).

\begin{figure}[t]
\begin{picture}(0,0)(150,255)   

\includegraphics{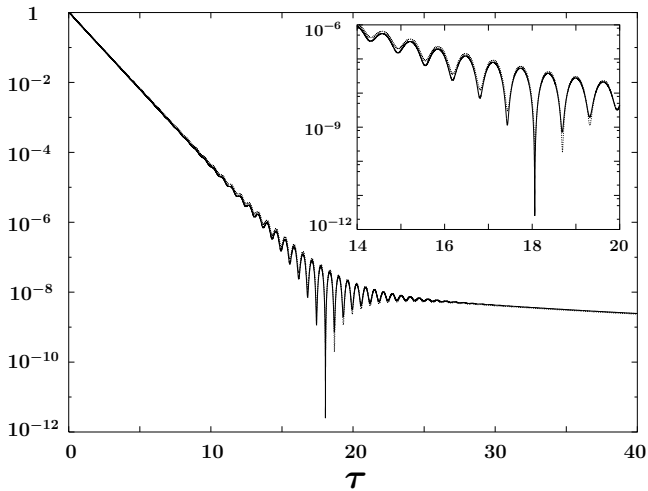}
\end{picture}
\vspace{6.4cm}
\caption{The probability $|c_e(\tau)|^2 = |\tilde{c}_e(\tau)|^2$ as a function of $\tau \equiv \Gamma_0 t$.
         The solid  curve corresponds to the exact numerical solution, i.e. solution of \eq{cp_RDL} with
	 the kernel  ${\tilde \kappa}^R_0(t)$ as in \eq{eq:kappa0} for a two-level system in vacuum.
	 The almost overlapping dotted curve corresponds to \eq{c_large_exp_corr}.
	 The values of the relevant parameters are  $\tilde{b}_A \equiv {\tilde{\omega}}_A/\Gamma_0=10$ and $\Lambda=1000$.
         The inset shows a limited part of the larger figure and illustrates the accuracy of the asymptotic form \eq{c_large_exp_corr}.
}
\label{c_large_tau_fig}
\end{figure}

    Finally, for large times as compared to the shifted atomic frequency transition, i.e. $\tilde{b}_A \tau \gg 1$,
    and also $({\tilde \Lambda} -1) \tilde{b}_A \tau \gg 1$, the kernel \eq{kappa_cisi_R_0} is, to first leading order 
\bea 
\label{kappa_approx}
  \tilde{\kappa}^R_0(\tau) \approx - \, \frac{\Gamma_0}{2}   +     \frac{\Gamma_0}{2\pi} \, \frac{ e^{i \tilde{b}_A \tau } }{ \tilde{b}_A \tau } \, .
\eea
   \noi
  One may then show that (see next section and Appendix \ref{B_app})
%
%
\bea  \label{c_large_exp_corr}
     \tilde{c}_e(\tau)   &\approx&   \, e^{-\tau/2} + \frac{ e^{i \tilde{b}_A \tau} }{2\pi i  \tau \bigg (  \tilde{b}_A - \ln (\tilde{b}_A\tau)/2\pi \bigg )^2}  \, ,
\eea
provided $\tilde{b}_A$ is sufficiently large, and where we have made use of ${\tilde \Lambda} \approx \Lambda$.
    \noi
    This solution includes the well-known exponential decay as well as a correction term which  dominates $|\tilde{c}_e(\tau)|^2$ for large values of $\tau$.
    \eq{c_large_exp_corr} is plotted in \fig{c_large_tau_fig} (dotted line)
    together with the exact numerical solution (solid line).
    The last term in \eq{c_large_exp_corr} will dominate over the exponential for times $\tau \gtrsim \tau^*$, where $\tau^*$
    is determined by the transcendental equation
\bea
\label{eq:tlnt}
  e^{-\tau^*/2} \approx \frac{ 1 }{2\pi \tau^* \bigg (  \tilde{b}_A -\ln(\tilde{b}_A \tau^* )/2\pi \bigg )^2}  \,  ,
\eea
valid for a sufficiently large value of $\tilde{b}_A$.
  For $\tilde{b}_A=10$ and $\Lambda = 1000$  as in \fig{c_large_tau_fig}, the solution of this equation is, e.g.,  $\tau^* \approx 18.5$. The $\tau \ln^2 (\tilde{b}_A\tau)$  dependence in Eqs.(\ref{c_large_exp_corr}) and (\ref{eq:tlnt})  becomes more important than the $\tau$ dependence only for very large time-scales $\tau \gtrsim \tau_{\ln}$, where $\tau_{\ln}= \exp(2\pi {\tilde b}_A)/{\tilde b}_A$, in which case $|{\tilde c}_e(\tau)|^2$ becomes exponentially close to zero with increasing value of ${\tilde b}_A$.

\vspace{0.5cm}
\bc{
\subsection{Asymptotic Expansion}
\label{sec_laplace_0}}
\ec
Laplace transform techniques have been used to investigate possible deviations from exponential decay as in Refs.\cite{knight_76,seke_88,seke_89}. As shown in particular by Seke and Herfort \cite{seke_88,seke_89}, careful considerations of the analytical properties of the decay amplitude are required in order to extract the asymptotic behavior of $c_e(t)$. The result of Refs.\cite{knight_76,seke_88,seke_89} may seem to be partly contradictory. It is therefore of some interest to investigate in detail in what manner the asymptotic behavior according to Eq.(\ref{c_large_exp_corr}) is obtained by making use of Laplace transform techniques. As we will see, it then becomes apparent that the various asymptotic forms of the decay amplitude are all valid but depends on the time-scale and other physical parameters at hand.

In order to obtain the corresponding asymptotic expansion, as $\tau$ becomes large, we therefore find it convenient to  consider the Laplace transform of the differential equation for the rescaled amplitude $\tilde{c}_e(t)$ as defined in
Eq.(\ref{trans}), i.e.
\bea   \label{ce_tilde}
  \frac{d\tilde{c}_e(t)}{dt} &=&  \int_0^t  dt^{\, \prime} ~ {\tilde K}^R_0(t-t^{\, \prime}) \, \tilde{c}_e(t^{\, \prime})\nonumber \\ &-&i\frac{\Gamma_0}{2\pi}\ln(\Lambda -1)\tilde{c}_e(t) \, ,
\eea
    \noi
    where the regularized kernel $K^R_n(t)$ is
\bea
\label{K_tilde}   
{\tilde K}^R_n(t)  =    - \, \frac{\Gamma_0}{2 \, \pi} \;  
                                     \int_0^{\omega_c} d\omega \,(\frac{\omega}{\omega_A})^n 
                                     e^{- \, i  (\omega - {\tilde \omega_A})  t  } \, ,\nonumber \\				    
\eea
with, for the moment, $n=0$. The Laplace transform $\tilde{c}_e(s)$ is then given by
\bea
\label{Laplace_n0}   
 \tilde{c}_e(s)=\frac{1}{s-{\tilde K}^R_0(s)+i\Gamma_0\ln(\Lambda -1)/2 \pi}\, , 
\eea
where we have identified
\bea
\label{eq:log}
{\tilde K}^R_0(s)=i\frac{\Gamma_0}{2\pi}\int_{0}^{\omega_c}d\omega \frac{1}{\omega- {\tilde \omega}_A -is}=~~~~~~~~~~~~\nonumber \\ 
i\frac{\Gamma_0}{2\pi} \bigg ( \ln(1- \frac{\omega_c}{is+{\tilde \omega}_A}) \bigg )\,,~~~~~~~~~~~~~~~~
\eea
in terms of the principal branch of the natural logarithm $\ln$-function.  The inverse Laplace transform in terms of the Bromwich integral and the dimensionless time parameter $\tau$ can then formally, after a suitable Wick ($u={\tilde b}_A+is/\Gamma_0$) rotation, be written in the form 
\bea
\label{eq:invlaplace}
\tilde{c}_e(\tau) = \frac{e^{i {\tilde b}_A\tau}}{2\pi i}\int_{\infty + i\gamma}^{-\infty +i\gamma}du e^{-i
u\tau}\times  ~~~~~~~~\nonumber \\
\left (  u-b_A  
 + \frac{1}{2\pi} 
 \ln(1-\frac{\Lambda_c}{u} )  \right)^{-1},~~
\eea
where we have defined
\bea
\label{eq:lambda_c}
\Lambda_c \equiv \frac{\omega_c}{\Gamma_0} \, ,
\eea
and where the positive real number $\gamma$ is chosen in such a way that the possible singularities of the integrand are above the integration contour.

We now follow the methods of, in particular, Refs.\cite{seke_88,seke_89}, where more details can be found, and use Eq.(\ref{eq:invlaplace}) in order to extract the large $\tau$, i.e.  $u\rightarrow 0$,  asymptotic expansion for $\tilde{c}_e(\tau)$. As in Refs.\cite{seke_88,seke_89}, we identify the functions
\bea
\label{eq:M0def}
M_0(u)\equiv u- b_A + \frac{1}{2\pi} [\log(u-\Lambda_c)- \log(u)] ,
\eea
%
and 
\bea
\label{eq:Mdef}
M_1(u)\equiv M_0(u)+ i ,
\eea
where the $\log$-function now stands for the multi-valued natural logarithm function.   The appropriate Riemann surface for the function $\log(u-\Lambda_c)- \log(u)$ has been obtained in Ref.\cite{seke_88}. Here one joins a second Riemann sheet at the branch-cut along the real $u$-axis from $u=0$ to $u =\Lambda_c$. 

The integration contour in Eq.(\ref{eq:invlaplace}) can then be deformed into  two curves of integration that runs parallel to the $\mbox{Im}(u)$-axis. One of these curves runs  from $-i\infty + \Lambda_c + \epsilon$ , above and around the branch point $u=\Lambda_c$ on the first Riemann sheet, and then towards $-i\infty +\Lambda_c -\epsilon$  on the second Riemann sheet, where $\epsilon$ is a small and positive real number. 
Similarly, the other curve of integration runs from $-i\infty + \epsilon$ on the second Riemann sheet, above and around the branch point $u=0$ on the first Riemann sheet, and then towards $-i\infty  -\epsilon$  on the first Riemann sheet.

The function $M_1(u)$ is obtained from $M_0(u)$ when passing to the second sheet of the Riemann surface construction. It can be shown that each of the functions $M_0(u)$ and $M_1(u)$ have no poles and only one zero on each of the Riemann sheets considered \cite{seke_88,seke_89}. On the first Riemann sheet, where the original functions in Eq.(\ref{eq:invlaplace}) are defined, there is, in addition, a zero of $M_0(u)$ along  the real axis at $u_0 \approx \Lambda_c(1+ \exp[-2\pi \Lambda_c])$. The corresponding residue is exponentially suppressed, i.e. $Z_0 \equiv 1/(dM_0(u)/du)_{u=u_0} \approx 2\pi\Lambda_c\exp(-2\pi\Lambda_c)$, apart from a phase factor, and the corresponding pole contribution has therefore been neglected.

In the deformation of the integration contour in Eq.(\ref{eq:invlaplace}), and on the second Riemann sheet, one now encounters  the conventional  Wigner-Weisskopf like 
pole-contribution    at $u=u_1 \approx \tilde{b}_A-i/2$, provided $\tilde{b}_A\gg 1$ due to Eq.(\ref{eq:Mdef}). 
Since $dM_1(u)/du = 1+\Lambda_c/2\pi u(u-\Lambda_c)\approx 1-1/{\tilde b}_A \approx 1$ for $u=u_1$ provided $\tilde{b}_A\gg 1$, this pole leads to a residue $Z_1 \equiv 1/(dM_1(u)/du)_{u=u_1} \approx 1$.  This Wigner-Weisskopf like 
pole-contribution therefore contributes  to  the amplitude $\tilde{c}_e(\tau)$ with $\exp(-\tau/2)$, i.e.  the expected and conventional amplitude describing exponential decay.  Including the two integrals  emerging from the branch points mentioned above we then, finally, obtain that the result
\bea
\label{eq:contours}
\tilde{c}_e(\tau) = e^{-\tau/2 } - \frac{e^{i {\tilde b}_A\tau}}{2\pi \tau}I_1 -
\frac{e^{i {\tilde b}_A(1-{\tilde \Lambda })\tau}}{2\pi \tau}I_2 \,,
\eea
where we have defined the integrals 
\bea
\label{eq:I1}
I_1 = \int_0^{\infty} dse^{-s}\left[\frac{1}{M_0(-is/\tau)}- \frac{1}{M_1(-is/\tau)}\right]\,,
\eea
and 
\bea
\label{eq:I2}
I_2 = ~~~~~~~~~~~~~~~~~~~~~~~\nonumber \\ \int_0^{\infty} dse^{-s}\left[\frac{1}{M_1(\Lambda_c-is/\tau)} - \frac{1}{M_0(\Lambda_c-is/\tau)}\right]\, ,
\eea
similar to the results of Refs.\cite{seke_88,seke_89}.
In Appendix \ref{B_app} we now show that, as $\tau \gg 1$,
\bea
\label{eq:I1asymp}
I_1 \approx \frac{i}{(\tilde {b}_A - \ln(\tilde {b}_A\tau)/2\pi)^2}  ,
\eea
provided $|b_A - \ln(\Lambda_c\tau)/2\pi|\approx |\tilde {b}_A - \ln(\tilde {b}_A\tau)/2\pi| \gg 1$. In the same manner it also follows that 
\bea
\label{eq:I2asymp}
I_2 \approx \frac{i}{(\Lambda_c - b_A -\ln(\Lambda_c\tau)/2\pi)^2}  ,
\eea
provided $ |\Lambda_c - b_A -\ln(\Lambda_c\tau)/2\pi| \gg 1$. As long as $\Lambda_c \gg b_A$ it is now clear that $I_1$ will dominate over $I_2$. Due to Eq.(\ref{eq:I1asymp}) we now recover the result as given in Eq.(\ref{c_large_exp_corr}). It now  also follows that $I_1 \approx 1/\tilde{b}_A^2$ unless $\tau$ is exponentially large, i.e. $\tau \gg e^{2\pi\tilde{b}_A}/\tilde{b}_A$, in which case the decay amplitude $\tilde{c}_e(\tau)$ is exponentially small. 

When compared to the analysis of Refs.\cite{seke_88,seke_89} we remark that our approach to the asymptotic behavior of integrals like Eqs. (\ref{eq:I1}) and (\ref{eq:app_I1}) has  a much broader range of applicability. In particular we notice that, as long as $\tau \ll e^{2\pi{\tilde b}_A}/{\tilde b}_A)$, the $I_1$ integral leads to a ${\cal O}(1/\tau)$ behavior of the decay amplitude ${\tilde c}_e(\tau)$. It is only in the opposite limit that the proposed ${\cal O}(1/\tau \ln^2\tau)$ of Refs.\cite{seke_88,seke_89} emerges, i.e. for $\tau \gg e^{2\pi{\tilde b}_A}/{\tilde b}_A$.  In this case the decay amplitude, however, becomes exponentially small for ${\tilde b}_A$ sufficiently large. We also find it important to express the asymptotic behavior in terms of dimensionless quantities like $\tau$ and not in terms of the  time variable $t$ with a physical dimension as in Refs.\cite{seke_88,seke_89}.

\bc{
\subsection{Comparison with Refs.\cite{knight_76,seke_88,seke_89}}
\label{sec_vacuum_knight_seke}}
\ec

%
 \begin{figure}[t]
\begin{picture}(0,0)(150,255)   

\includegraphics{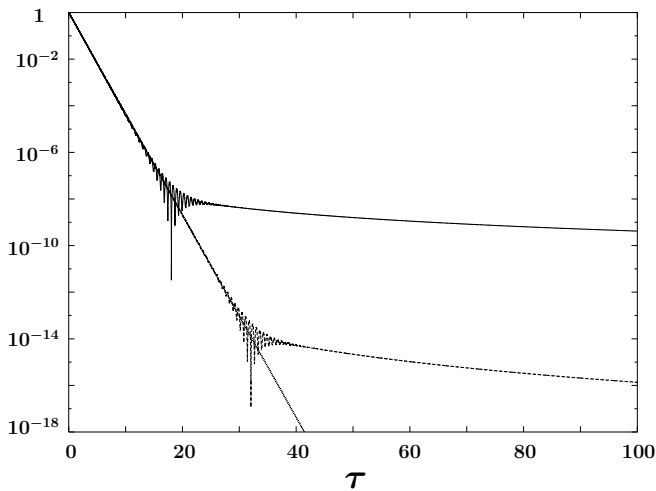}
\end{picture}
\vspace{6.4cm}
\caption{The probability $|c_e(\tau)|^2=|{\tilde c}_e(\tau)|^2$ as a function of $\tau \equiv \Gamma_0 t$.
        The upper curve corresponds to the exact numerical solution, i.e. solution of \eq{cp_RDL} with
	 the kernel ${\tilde \kappa}_0^R(t)$ as in \eq{kappa_cisi_R_0} for a two-level system in vacuum which, within the numerical accuracy of the figure, overlaps with the asymptotic formula Eq.(\ref{c_large_exp_corr}).
	 The lower curve corresponds to the main result in Ref.\cite{knight_76}, i.e. \eq{Eq_9_PLK}. The asymptotic form according  \eq{PLK_asymp} is in excellent agreement with the numerical evaluation of \eq{Eq_9_PLK}.
	 The values of the relevant parameters are  ${\tilde b}_A =10$ and $\Lambda=1000$. As a reference, the straight line in the Figure corresponds to a pure exponential decay, i.e. $|c_e(\tau)|^2=e^{-\tau}$.
}
\label{large_tau_fig}
\end{figure}

   As alluded to above, in many cases of the study of deviations from  the exponential decay of  atomic spontaneous emission processes, various approximations have been applied in order to obtain explicit and analytical expressions 
as, e.g.,  in Refs.\cite{knight_76,seke_88,seke_89} and references cited therein. In these references one consider electric- dipole transitions and,  in the integral equation \eq{cp_DL} for the vacuum case, the kernel $\kappa (\tau)$ should therefore be replaced by kernel $\kappa_1^R (\tau)\equiv \kappa_1 (\tau)$
as defined by  \eq{kappa_ins}, i.e. with a factor $\omega/\omega_A$ in the integrand rather than $(\omega/\omega_A)^3$. Contrary to the regularization procedure  employed in Section  \ref{sec_vacuum}, this factor $\omega/\omega_A$ is, however, kept untouched and a suitable subtraction is carried out at a later stage in the analysis.   As we now will verify the actual subtraction procedure in order to make the corresponding kernel finite  will effect the large time deviation from exponential decay. If the cut-off frequency $\omega_c$ again is regarded to  be a finite physical quantity, we notice that 
   the regularized kernel $\kappa^{R}_1(\tau)$ may be written in terms of $\kappa^R_0(\tau)$ as given by \eq{kappa_ins_R}:
\bea  \label{eta_tilde}
   \kappa_1^R(\tau)  = ~~~~~~~~~~~~~~~~~~~~ \nonumber \\
 \kappa^R_0(\tau)   + 
						      \frac{\Gamma_0}{{2\pi b_A }  \tau} \bigg ( e^{-i (\Lambda-1) b_A \tau} - e^{i b_A \tau} \bigg )
	                                              +  
					              i \, \frac{\Gamma_0}{2\pi} \, \Lambda \, .
\eea
    \noi
    Due to this identity we realize that the kernel $\kappa_1^R(\tau)$ is linear in the dimensionless cut-off
    frequency rather than the logarithmic dependence of $\kappa^R_0(\tau)$ as in \eq{kappa_temp}. We observe that the linear $\Lambda$-term may now be absorbed in an energy shift such that the kernel $\kappa_1^R(\tau)$ is replaced according to:
\bea  \label{eta_tilde_ls}
   \kappa_1^R(\tau)  \rightarrow    \kappa_1^R(\tau) = ~~~~~~~~~~~~~~ \nonumber \\
\kappa^R_0(\tau)  +   \frac{\Gamma_0}{{2\pi \, b'_A }  \tau} \, \bigg ( e^{-i (\Lambda-1) b'_A \tau} - e^{i b'_A \tau} \bigg ) \, , 
\eea
     \noi
     where $b'_A \equiv \omega'_A/\Gamma_0$, and
\bea   \label{omega_A_prime}
    \omega'_A \equiv \omega_A - \frac{\Gamma_0}{2\pi} \Lambda \; ,
\eea
and where also the $\omega_A $-dependence in $\kappa^R_0(\tau)$ is replaced by $\omega'_A$. Such a frequency shift preserves unitarity.  As in Refs.\cite{knight_76,seke_88,seke_89}, the linear $\Lambda$-dependence can therefore be removed altogether by applying Bethe's mass renormalization \cite{bethe_47}, i.e. by including a suitable mass counter-term in the Hamiltonian. After this removal of the linear $\Lambda$-dependence and by, in addition, applying the transformation in \eq{trans}, the kernel $\kappa^R_1(\tau)$ is transformed into the kernel $\tilde{\kappa}^R_1(\tau)$ according to:
\bea  \label{eta_tilde_ls}
   \kappa^R_1(\tau)  \rightarrow    \tilde{\kappa}^R_1(\tau) = ~~~~~~~~~~~~~~ \nonumber \\  \tilde{\kappa}^R_0(\tau)  +   \frac{\Gamma_0}{{2\pi \, \tilde{b}_A }  \tau} \, \bigg ( e^{-i (\tilde{\Lambda}-1) \tilde{b}_A \tau} - e^{i \tilde{b}_A \tau} \bigg ) \, .
\eea
The asymptotic form of the kernel $\tilde{\kappa}^R_1(\tau)$ is now such that the sub-leading contribution from the asymptotic from of the kernel $\tilde{\kappa}^R_0(\tau)$, as given by Eq.(\ref{kappa_approx}), is cancelled and therefore
\bea 
\label{eta_approx}
  \tilde{\kappa}^R_1(\tau) \approx - \, \frac{\Gamma_0}{2}   +     \frac{\Gamma_0}{2\pi} \, \frac{ e^{-i (\tilde{\Lambda}-1) \tilde{b}_A \tau} }{ \tilde{b}_A \tau } \, ,
\eea
for sufficiently large $\tau$.  Hence, the methods of Appendix \ref {A_app} would then lead to
\bea  \label{c_eta_large_exp_corr}
     \tilde{c}_e(\tau)   &\approx&   \, e^{-\tau/2} - \frac{ e^{-i(\tilde{\Lambda}-1) \tilde{b}_A \tau} }{2\pi i  \tau \tilde{\Lambda}\tilde{b}_A^2}  \, ,
\eea
apart from possible $1/\tau^2$ corrections provided $\tilde{b}_A$ is sufficiently large. 

As in Section \ref{sec_laplace_0},  a more precise asymptotic expansion as $\tau$ becomes large, and which also takes the Lamb shift Eq.(\ref{omega_A_tilde}) more carefully into account, can be obtained by a consideration of a Bromwich integral.  The Laplace transform of the differential equation for the rescaled amplitude $\tilde{c}_e(t)$ reads 
%
%
\bea   \label{ce1_tilde}
  \frac{d\tilde{c}_e(t)}{dt} &=&  \int_0^t  dt^{\, \prime} ~ {\tilde K}^R_1(t-t^{\, \prime}) \, \tilde{c}_e(t^{\, \prime})\nonumber \\ &-&i\frac{\Gamma_0}{2\pi}\ln(\Lambda -1)\tilde{c}_e(t) - i\frac{\delta m \,c^2}{\hbar}\tilde{c}_e(t) \, ,
\eea
where the regularized kernel ${\tilde K}^R_1(t)$ is
\bea
\label{K1_tilde}   
{\tilde K}^R_1(t)  =    - \, \frac{\Gamma_0}{2 \, \pi} \;  
                                     \int_0^{\omega_c} d\omega \,\frac{\omega}{\omega_A}
                                     e^{- \, i  (\omega - {\tilde \omega}_A)  t  } \, ,\nonumber \\				    
\eea
and where the unitarity preserving transformation as defined by Eq.(\ref{trans}) has been applied.
%
%
%
%
In Eq.(\ref{ce1_tilde})  we have also included the contribution due to the additional mass counter-term $\delta m \,c^2/\hbar$ in the Hamiltonian which has the role of cancelling the linear $\Lambda$-dependence in the kernel Eq.(\ref{K1_tilde}). 
After a tedious but straightforward analysis, following the same Laplace transformation techniques as in Refs.\cite{seke_88,seke_89}, the large $\tau$ expansion of $\tilde{c}_e(\tau)$ can now be obtained. Similar to the definitions of $M_0(u)$ and $M_1(u)$ in Section \ref{sec_laplace_0}, we therefore introduce the functions
\bea
\label{eq:N0def}
N_0(u)\equiv u- b_A + \frac{u}{2\pi b_A} \big(\log(u-\Lambda_c)- \log u \big) ,
\eea
%
and 
\bea
\label{eq:N1def}
N_1(u)\equiv N_0(u)+ i\frac{u}{b_A}\, ,
\eea
where the linear $\Lambda$-dependence in Eq.(\ref{ce1_tilde}) now has explicitly been removed by the mass renormalization procedure. The solution to the Eq.(\ref{ce1_tilde}) can then be written in the form
\bea
\label{eq:invlaplace_2}
\tilde{c}_e(\tau) = \frac{e^{i {\tilde b}_A\tau}}{2\pi i}\int_{\infty + i\gamma}^{-\infty +i\gamma}du e^{-iu\tau}\frac{1}{N_0(u)} \, .
\eea
By deforming the integration contour in Eq.(\ref{eq:invlaplace_2}) in the same manner as in the derivation of Eq.(\ref{eq:contours}), one then obtains the result
\bea
\label{eq:contours1}
\tilde{c}_e(\tau) = e^{-\tau/2 } - \frac{e^{i {\tilde b}_A\tau}}{2\pi \tau}J_1 -
\frac{e^{i {\tilde b}_A(1-{\tilde \Lambda })\tau}}{2\pi \tau}J_2 \,,
\eea
where we have defined the integrals 
\bea
\label{eq:J1}
J_1 = \int_0^{\infty} dse^{-s}\bigg( \frac{1}{N_0(-is/\tau)}- \frac{1}{N_1(-is/\tau)}\bigg)\,,
\eea
and 
\bea
\label{eq:J2}
J_2 = ~~~~~~~~~~~~~~~~~~~~~~~\nonumber \\ \int_0^{\infty} dse^{-s}\bigg( \frac{1}{N_1(\Lambda_c-is/\tau)} - \frac{1}{N_0(\Lambda_c-is/\tau)}\bigg)\, .
\eea
We have therefore actually reproduced the results of Refs.\cite{seke_88,seke_89}, but now expressed in terms of dimensionless parameters. 

As expanded upon in the Appendix \ref{C_app}, the  integrals $J_1$ and $J_2$ can now be analyzed in manner similar to the integrals $I_1$ and $I_1$ of Section \ref{sec_laplace_0} .
     If $\tau \gg 1$ and $\Lambda_c \gg 1$ and, in  addition,  $b_A\tau \gtrsim 1 + \ln (\Lambda_c)/2\pi b_A$ ,  we then obtain
\bea   \label{C}
    J_1 \approx \frac{1}{b_A^3\tau}\, .
\eea
   \noi
For the integral $J_2$ it also follows that, for $\tau \gg 1$, 
\bea   \label{C}
    J_2 \approx \frac{1}{i\Lambda (b_A -\ln (\Lambda_c \tau)/2\pi)^2}\, .
\eea
 provided $|b_A -\ln (\Lambda_c \tau)/2\pi| \gg 1$.  The insertion of these asymptotic expansions into Eq.(\ref{eq:contours1}) therefore leads to
\bea   \nonumber
\label{eq:general_c}
    \tilde{c}_e(\tau) ~ \approx ~  e^{-\tau/2} ~ - ~ e^{i \tilde{b}_A \tau} \, \frac{1}{2\pi \, b_A^3} \, \frac{1}{\tau^2}  
                       \\ \nonumber
		       \\ \label{c_asymp}
		       -  \frac{e^{- i \, (\tilde{\Lambda}-1) \, \tilde{b}_A \tau}}{2\pi \, i \,  \Lambda} \; \frac{1}{\tau}  ~
		            \frac{1}{( \, b_A  - \ln( \Lambda_c \, \tau )/2\pi \, )^2} \, .
\eea 
The last term in Eq.(\ref{eq:general_c}) with the  $(b_A - \ln( \Lambda_c \, \tau )/2\pi)$-dependence has been obtained by using the technique of the Appendix \ref{B_app} in order to see how the claimed $\tau\ln^2 \tau$ dependence of Ref.\cite{seke_88,seke_89} actually emerges.  Here we again stress that our results are expressed in terms of dimensionless quantities in contrast to the results of Refs.\cite{seke_88,seke_89}.

   At this point we now quote the  following explicit, but approximative, expression for the probability amplitude $c_e(\tau)$ result of  Knight and Milonni in Ref.\cite{knight_76}:
\bea   \nonumber
&&  c_e(\tau) =  e^{-  \tau/2 + i\delta b\tau}  - \frac{e^{i b_A \tau}}{2\pi \, i } \, 
                 \\ \nonumber
		 && \times \int_0^{\infty} dx  \, e^{- \Lambda \, x \, b_A\tau } \bigg (  
                 \frac{1}{x -  \frac{i}{\Lambda} + i \, \frac{x}{2b_A}  - \frac{x}{2\pi \, b_A} \, [ \ln x - i \, \frac{\pi}{2} ] } 
	         \\  \label{Eq_9_PLK}
	         && ~~~~~~~~~~~~\, - ~ 
	         \frac{1 }{x -  \frac{i}{\Lambda} -  \frac{x}{2\pi \, b_A} \, [ \ln x + i \, \frac{\pi}{2} ] }  ~ \bigg )\, ,  ~~~~~~~~~~~~~  
\eea
   \noi
   where  $\delta b = \delta \omega/\Gamma_0$,  and
\bea
   \delta \omega = \frac{\Gamma_0}{2\pi} \, \ln (\Lambda -1)\, ,
\eea
  \noi
  is the angular frequency  shift of the excited state $|e \rangle$. This expression, which we have verified  \cite{comment_knight}, is obtained by deforming integration contours of the Bromwich integral by only considering the analytical properties of the natural logarithm ln-function and not taking into account the additional analytical conditions imposed by the appearance of the functions $N_0(u)$ and $N_1(u)$ as discussed  above. 
  For large times, i.e. $b_A\tau \gg 1$, this expression is reduced to \cite{knight_76,comment_knight}
\bea   \label{PLK_asymp}
   c_e(\tau)  &\approx&  e^{-  \tau/2 + i\delta b \tau} \, - \, \frac{1}{2\pi \, b_A} \, \frac{1}{(b_A\tau)^2} \, e^{i b_A \tau } \, .
\eea
   \noi
   We notice that up to a phase redefinition, this equation and Eq.(\ref{eq:general_c}) only differ by that last Seke-Herfort term  of Eq.(\ref{eq:general_c}).  It is only in the limit $\Lambda \rightarrow \infty$ that Eq.(\ref{eq:general_c}) leads to $1/\tau^2$ asymptotic behavior of Knight and Milonni. It is now also clear that it is only for time-scales $\tau \gtrsim e^{2\pi b_A}/\Lambda_c$ that the $1/\tau \ln^2 \tau$ asymptotic behavior of Seke-Herfort emerges. Such time-scales are typically exponentially large leading to an exceedingly small decay amplitude $c_e(\tau)$ unless $\Lambda_c$ is very large. But as we have seen above, in such a case the $1/\tau^2$ asymptotic behavior of Knight and Milonni begins to dominate. We also realize that Eq.(\ref{eq:general_c}) leads to a dominant $1/\tau$ behavior for $\tau \lesssim e^{2\pi b_A}/\Lambda_c$ unless $\Lambda_c$  again is not arbitrarily large. 

By a comparison of Eqs.(\ref{c_large_exp_corr}) and (\ref{eq:general_c}) we also realize, due to the presence of the factor $1/\Lambda$ in the Seke-Herfort term  in Eq.(\ref{eq:general_c}), that the regularization procedure actually effects the large-time behavior of the decay amplitude $c_e(\tau)$. 

\begin{figure}[t]
\begin{picture}(0,0)(150,255)   

\includegraphics{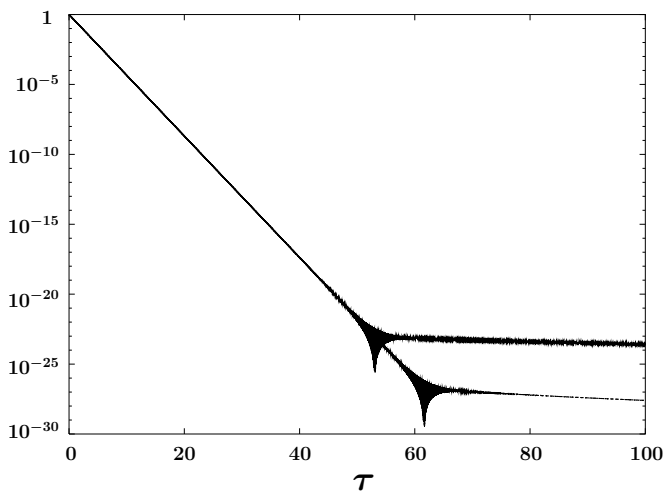}
\end{picture}
\vspace{6.4cm}
\caption{The probability $|c_e(\tau)|^2=|{\tilde c}_e(\tau)|^2$ as a function of $\tau \equiv \Gamma_0 t$.
         The upper curve corresponds to the exact numerical solution of the Volterra equation as considered in Ref.\cite{seke_88,seke_89}, i.e.  of \eq{cp_RDL} with
	 the kernel ${\tilde \kappa}^R_1(\tau)$ of \eq{eta_tilde_ls} for a two-level system in vacuum. For large values of $\tau$ the solution is dominated by the $1/\tau$ asymptotic behavior according to Eq.(\ref{eq:general_c}). The lower curve corresponds to Eqs. (\ref{Eq_9_PLK}) and (\ref{PLK_asymp}), which overlaps within the numerical accuracy of the figure, with a characteristic $1/\tau^2$ asymptotic behavior for $c_e(\tau)$.
	 The values of the relevant parameters are $\tilde{b}_A =1000$ and $\Lambda=1000$.
}
\label{seke_fig}
\end{figure}

   We illustrate some of these features in \fig{large_tau_fig} and \fig{seke_fig}. For comparison, Eqs.(\ref{Eq_9_PLK}) and (\ref{PLK_asymp}) are plotted in \fig{large_tau_fig} together with the exact numerical solution.
In \fig{seke_fig} we have plotted the solution of \eq{cp_RDL} with the kernel ${\tilde \kappa}^R_1(\tau)$ of
  \eq{eta_tilde_ls} together with  Eqs. (\ref{Eq_9_PLK}) and (\ref{PLK_asymp}). As long as the Seke-Herfort term in Eq.(\ref{eq:general_c}) dominates for sufficiently large values of $\tau$, this contribution will dominate the conventional exponential decay term if $\tau \gtrsim \tau_{*}$, where $\tau_{*}$ is determined by the transcendental equation
\bea
\label{eq:sh_tlnt}
  e^{-\tau_{*}/2} \approx \frac{ 1 }{2\pi \Lambda\tau_{*} \bigg ( b_A -\ln(\Lambda_c \tau_{*} )/2\pi \bigg )^2}  \,  ,
\eea
With the choice of parameters as in  \fig{seke_fig} and $\tau \gtrsim \tau_{*} \approx 53$, one finds that Eq.(\ref{eq:general_c}) in this case leads to $1/\tau$ scaling rather than the $1/\tau^2$ asymptotic scaling for the decay 
amplitude $c_e(\tau)$ according to \eq{Eq_9_PLK}.

\vspace{0.8cm}

\bc{
\section{FINAL REMARKS}
\label{sec:summary}}
\ec
%
%

    We have seen that the procedure of regularization the decay amplitude $c_e(\tau)$ actually influences the deviation from exponential decay at large values of the time-parameter $\tau$.  It would, of course, be of interest if one e.g. in terms of artificial atoms could experimentally investigate deviations from exponential decay at large times and thereby infer a possible presence of the Seke-Herfort term of Eq.(\ref{eq:general_c}) or the related $1/\tau$ behavior  obtained in Eq.(\ref{c_large_exp_corr}). As alluded to in the introductory remarks  and in the case of atom chips,  the presence of material bodies can drastically change the characteristic lifetime (see e.g. Refs. \cite{henkel_99,rekdal_04, skagerstam_06, rekdal_07,hinds_2007}) and one could then perhaps expect to encounter deviations from exponential decay. With the parameters of Refs.\cite{rekdal_04,rekdal_07,hinds_2007}, i.e. $S^2 =1/8$, $g_S \approx 2$ and $\omega_A/2\pi = 560$ kHz, we have that $b_A\approx 3\cdot 10^{32}$.   Using the cut-off $\Lambda$ according to Eq.(\ref{Lambda_cut_off}), i.e. $\Lambda \approx 2\cdot 10^{14}$ one finds that $b_A\approx {\tilde b}_A$ and that $\tau^* \approx 315$ and $\tau_* \approx 380$ according to Eqs.(\ref{eq:tlnt}) and (\ref{eq:sh_tlnt}) respectively. 
Since our analysis involves the dimensionless parameter $\tau$ this means that even though the effective decay-rate  $\Gamma$ can be changed by, say, twenty orders of magnitude (see e.g. Refs. \cite{henkel_99,rekdal_04, skagerstam_06, rekdal_07,hinds_2007}), one still would have to consider several hundred decay times in order to see deviations from exponential decay.  
This means that for atom chips the conventional exponential decay description applies with an exceedingly high accuracy. As we, however, have remarked in the introduction,  experimental studies have e.g.  shown that artificial atoms can lead to a Lamb shift of the order of a few per cent of the typical emission line \cite{fragner_2008}. We now estimate the corresponding relevant parameters according to Ref.\cite{fragner_2008}. The natural frequency is $\omega _A/2\pi \approx {\cal O}$(GHz) and the effective rate $\Gamma_0$ is estimated from the Lorentzian fit to the experimental data to be
$\Gamma_0 \approx 0.5\cdot 10^{-2} \omega _A/2\pi$. With a Lamb-shift of the order of one per cent of $\omega_A$, we then have that $\Gamma_0 \ln (\Lambda -1)/2\pi \approx 10^{-2}\omega_A$, i.e. the effective cut-off $\Lambda$ is then given by $\Lambda \approx e^{8\pi^2}$. One then finds  somewhat more optimistic numbers $\tau^* \approx 40$ and $\tau_* \approx 200$ according to Eqs.(\ref{eq:tlnt}) and (\ref{eq:sh_tlnt}), respectively, which may open the door for observing deviations from exponential decay in system composed of artificial atoms.


\begin{center}ACKNOWLEDGEMENTS
\end{center}

\vspace{0.2cm}

     This work has been supported in part by the Norwegian University of Science and Technology (NTNU)
     and for one of the authors (BSS) also by the Norwegian Research Council under contract: NFR 191564/V30~"Complex Systems and Soft Materials". BSS acknowledge discussions over the years with members of the TH-division at CERN on issues related to the present research topic and  the hospitality shown at CAS during the final preparation of the present paper. We are grateful to prof. J. Wilkening and  R.\,Rekdal for various suggestions in the numerical work. One of the authors (PKR) also gratefully acknowledges financial support from Kalamaris Invest AS and F. O. F. Johannessen.

\vspace{0.5cm}

\appendix

\section{Large time expansion}
\label{A_app}

   In section \ref{sec_vacuum_approx}, the kernel $\tilde{\kappa}_0^R(\tau)$ was considered for three different regimes: small, intermediate and large times $\tau$ for a given $\tilde{b}_A$ and given $\Lambda$. The integral in the Volterra equation \eq{cp_RDL} may then trivially be split into these three
   intervals, where one interval relieves the other. For large times $\tau \gg 1$ and, in addition $\tilde{b}_A \gg 1$, two of these integrals are
   negligible. The Volterra equation in \eq{cp_RDL} is then left with an integral corresponding to the kernel in
 \eq{kappa_approx}.   This integral may be computed analytically:
\bea
\label{eq:c_asymptotic}
\tilde{c}_e(\tau) - 1  \approx  \int_0^{\tau-1/\tilde{b}_A}dx \bigg\{-\frac{1}{2} +  \frac{1}{2\pi} \frac{e^{ i \, \tilde{b}_A (\tau-x)}}{\tilde{b}_A (\tau-x)} \,\bigg\}e^{-x/2} \nonumber \\
	= -1 + e^{-\tau/2 + 1/2\tilde{b}_A} -\frac{1}{2\pi \tilde{b}_A}E_1[-(1/2 + i \tilde{b}_A) \tau ]e^{-\tau/2} \nonumber \\
	 +   \frac{1}{2\pi \tilde{b}_A}E_1[-(1/2\tilde{b}_A + i)] \, e^{-\tau/2}\, ,~~~
\nonumber \\
\eea
   \noi
   where $E_1(z)$ is the exponential integral for a complex argument $z$.
   As $\tau \gg 1$ and $\tilde{b}_A \gg 1$, the latter expression may now be simplified by making use of the leading asymptotic expansion of the $E_1$ functions in Eq.(\ref{eq:c_asymptotic}) for large values of the parameter $\tau$:
\bea
\label{eq:Asymp}
  \tilde{c}_e(\tau)  \approx   e^{-\tau/2}  +  \frac{e^{i \tilde{b}_A \tau}}{2\pi i \, \tilde{b}_A^2 \, \tau}\frac{1}{1-i/2\tilde{b}_A} ~~ , 
\eea

  \noi
  i.e. \eq{c_large_exp_corr} if $\tilde{b}_A \gg 1$.  As alluded to  in the main text Eq.(\ref{eq:Asymp}) remains valid for large values of $\tau$ provided $\tilde{b}_A \gg \ln(\tilde{b}_A\tau)/2\pi$, a condition that emerges from further iterations of the Volterra equation in \eq{cp_RDL}.  As discussed in Appendix \ref{B_app}, a more precise asymptotic form than the one given by Eq.(\ref{eq:Asymp}) can be obtained by making use of Laplace transform techniques.
\vspace{5mm}
\section{Asymptotic Expansion of $I_1$ and $I_2$}
\label{B_app}

Here we first consider the integral $I_1$ as defined in Eq.(\ref{eq:I1}) in Section  \ref{sec_laplace_0} which we rewrite in the following convenient form 
\bea
\label{eq:app_I1}
&&I_1 = \int_0^{\infty} dse^{-s}\left[\frac{1}{-is/\tau -b_A + \ln(1+ \Lambda_c \tau/is )/2\pi}- \right. \nonumber \\
&&~~~~~\left. \frac{1}{-is/\tau -b_A + \ln(1+ \Lambda_c \tau/is )/2\pi +i }\right] \,  ,
\eea
i.e., as $\tau$ becomes large,
\bea
\label{eq:app_I1_approx}
I_1 \approx  -\int_0^{\infty} dse^{-s}\left[\frac{1}{ a + \ln s/ 2\pi +i/4} - \right. \nonumber \\
\left. \frac{1}{a  + \ln s /2\pi -3i/4 }\right]\,,
\eea
where we have defined 
\bea
\label{def_a}
a\equiv b_A - \frac{\ln(\Lambda_c\tau)}{2\pi} \approx {\tilde b}_A - \frac{\ln ({\tilde b}_A \tau )}{2\pi} \, ,
\eea
 if $\Lambda \approx {\tilde \Lambda }$. One now finds, by a combination of analytical and numerical methods, that the real part of $I_1$ can be neglected, provided
$a$ is big enough, i.e. $|a| \gtrsim 10$,   and hence $I_1 \approx i/a^2$. By making use of the same reasoning as above one can now verify that $I_2 \approx i/{\tilde a}^2$, where now ${\tilde a} \equiv \Lambda_c - b_A -\ln(\Lambda_c\tau)/2\pi$ with $\Lambda_c \equiv \omega_c/\Gamma_0$.  Therefore the contribution due to $I_2$ can be neglected  in comparison with $I_1$, for any reasonable range of $\tau$, provided that $\Lambda_c \gg {\tilde b}_A$.
\vspace{3mm}
\section{Asymptotic Expansion of $J_1$ and $J_2$}
\label{C_app}

Here we first consider the integral $J_1$ as defined in Eq.(\ref{eq:J1}) in Section  \ref{sec_vacuum_knight_seke} for large values of $\tau$ where we, due to the presence of the factor $e^{-s}$ in the integrand of $J_1$,  make use of the approximations
\bea
N_0(-is/\tau) \approx -b_A\left( 1+ i\frac{s}{b_A \tau} \delta  \right)\, ,
\eea
with $\delta \equiv 1+\ln (\Lambda_c)/2\pi b_A$, 
as well as 
\bea
N_1(-is/\tau) \approx -b_A\left( 1+ i\frac{s}{b_A \tau} \delta  \right) +\frac{s}{b_A \tau} \, ,
\eea
as $\tau$ becomes large. A series expansion in $s/b_A\tau$ then leads to the convergent integral
\bea
J_1 \approx \frac{1}{b_A^3\tau } \int_0^\infty ds e^{-s}  \frac{s} {(1+is\delta/b_A \tau)^2} \, .
 \eea
The effective expansion time-parameter is then actually $b_A\tau$ instead of $\tau$. We therefore find  that $J_1  \approx 1/b_A^3\tau$ for sufficiently large values of the parameter $b_A\tau$ , i.e.  $b_A\tau \gtrsim 1+\ln (\Lambda_c)/2\pi b_A$. In the opposite limit $J_1$ will be suppressed by factors like $\ln (\delta/b_A\tau)/(\delta/b_A\tau)^2$ in addition to the $1/b_A^3\tau$ behavior. More care has to be taken when considering the integral $J_2$. For large values of $\tau$, we then make use of the expansions
\bea
N_0(\Lambda_c-is/\tau) \approx \Lambda\left( a -\frac{1}{4}i+\frac{\ln s}{2\pi} \right) \, ,
\eea
with $a$ as in Eq.(\ref{def_a}), 
as well as 
\bea
N_1(-is/\tau) \approx \Lambda \left( a + \frac{3}{4}i+\frac{\ln s}{2\pi} \right) \, .
\eea
%
%
%
%
Here we have made use of the simple fact that $\Lambda_c=\Lambda b_A$.
As long as $|a|$ is big enough it can now be argued that we can neglect $\ln s/2\pi$ term in the $N_0$ and $N_1$ functions above. This is so since for large $s \gtrsim e^{2\pi|a|}$ the integrand of $J_2$ is exponentially suppressed while for sufficiently small $s \lesssim e^{-2\pi|a|}$ there will only be an exponentially small range of integration. Proceeding in this manner and by series expansions in $-i/4a$  and $3i/4a$, we then find that the dominant contribution to $J_2$ is imaginary and that $J_2 \approx 1/i{\tilde \Lambda}a^2$. By numerical methods we find that  this is indeed an excellent approximation if $|a| \gtrsim 10$ when evaluating $|c_e(t)|^2$.


\begin{thebibliography}{xx}
\bibitem{comment_knight} We have corrected for some minor misprints in Eq.(9) of Ref.\cite{knight_76}.
\end{thebibliography}
\end{document}